# Correlating Stroke Risk with Non-Invasive Tracing of Brain Blood Dynamic via a Portable Speckle Contrast Optical Spectroscopy Laser Device

Yu Xi Huang,[a,†] Simon Mahler,[a,†,**] Aidin Abedi,[b,c,d] Julian Michael Tyszka,[e] Yu Tung Lo,[b,f] Patrick D. Lyden,[g] Jonathan Russin,[b,c] Charles Liu,[b,c,*,***] Changhuei Yang[a,***]

[a] Department of Electrical Engineering, California Institute of Technology; Pasadena, CA 91125, USA.
[b] USC Neurorestoration Center, Department of Neurological Surgery, Keck School of Medicine, University of Southern California; Los Angeles, CA 90033, USA.
[c] Rancho Research Institute, Rancho Los Amigos National Rehabilitation Center; Downey, CA 90242, USA.
[d] Department of Urology, University of Toledo College of Medicine and Life Sciences; Toledo, OH 43614, USA.
[e] Division of Humanities and Social Sciences, California Institute of Technology; Pasadena, CA 91125, USA.
[f] Department of Neurosurgery, National Neuroscience Institute, Singapore 308433
[g] Department of Physiology and Neuroscience, Zilkha Neurogenetic Institute, and Department of Neurology, Keck School of Medicine, University of Southern California, Los Angeles, CA 90033 USA
[†] These authors contributed equally to this work.
[***] These authors co-supervised this work.
**Corresponding authors:**
*Email: cliu@usc.edu
**Emails: mahler@caltech.edu, sim.mahler@gmail.com

**One Sentence Summary:** We present a non-invasive speckle contrast spectroscopy device for stroke risk pre-screening via cerebral blood flow analysis during breath-holding.

## Abstract

Stroke poses a significant global health threat, with millions affected annually, leading to substantial morbidity and mortality. Current stroke risk assessment for the general population relies on markers such as demographics, blood tests, and comorbidities. A minimally invasive, clinically scalable, and cost-effective way to directly measure cerebral blood flow presents an opportunity. This opportunity has potential to positively impact effective stroke risk assessment prevention and intervention. Physiological changes in the cerebral vascular system, particularly in response to carbon dioxide level changes and oxygen deprivation, such as during breath-holding, can offer insights into stroke risk assessment. However, existing methods for measuring cerebral perfusion reserve, such as blood flow and blood volume changes, are limited by either invasiveness or impracticality. Here, we propose a transcranial approach using speckle contrast optical spectroscopy (SCOS) to non-invasively monitor regional changes in brain blood flow and volume during breath-holding. Our study, conducted on 50 individuals classified into two groups (low-risk and higher-risk for stroke), shows significant differences in blood dynamic changes during breath-holding between the two groups, providing physiological insights for stroke risk assessment using a non-invasive quantification paradigm. Given its cost-effectiveness, scalability, portability, and simplicity, this laser-centric tool has significant potential in enhancing the pre-screening of stroke and mitigating strokes in the general population through early diagnosis and intervention.

## Introduction

Stroke is a global health concern, with a distressingly high incidence and substantial morbidity and mortality[1]. Annually, more than ten million people worldwide are impacted by strokes, imposing a heavy toll on affected individuals and their families, with significant health, financial, and quality of life burdens[2]. From the age of 25, the lifetime risk of stroke is around 25%[3, 4]. In the US, stroke affects nearly 800,000 individuals each year, contributing to significant healthcare challenges[5]. Early detection of stroke risk is pivotal in reducing its incidence and severity and improving its outcomes. A variety of interventions, such as lifestyle modifications (including diet, exercise, and smoking cessation) and medical management strategies (such as blood pressure control), exist to mitigate stroke risks[5–8].

Traditionally, stroke risk estimation relies on indirect markers such as demographics and comorbidities. As such, the most widely used questionnaires for stroke risk estimation capture factors such as diabetes and hypertension, along with lifestyle-related factors such as smoking and low physical activity[2, 9–14]. These risk profiles are based on population data and are useful for general assessments.

However, they are not definitive for determining the need for invasive or non-invasive evaluations, nor for guiding surgical or pharmacological interventions in individual patients to prevent future strokes. Limited non-invasive, scalable, and cost-effective measures based on direct neuroimaging currently exist to stratify stroke risk in the general population. This gap highlights the pressing need for objective and cost-effective measures of cerebral perfusion reserve, particularly evident when considering the management of myocardial ischemia through stress testing. Stress tests serve as a pivotal, objective method to evaluate cardiac health, functional capacity, and areas susceptible to ischemia. These tests leverage the shared pathophysiological pathways of cerebrovascular compromise and coronary artery disease. Yet, an equivalent direct and objective measure for assessing neurovascular health that is low-cost enough to deploy broadly in the community—akin to but potentially more available than cardiac stress testing—is conspicuously absent. We believe that this study introduces a simple, cost-effective solution that does not require ionizing radiation or radioactive isotopes that has the potential to lead to better development of personalized strategies for reducing stroke risk.

Efforts to characterize cerebral perfusion capacity have been underway for the past four decades. Neuroimaging modalities such as PET, SPECT, xenon-enhanced CT, and perfusion CT can be coupled with Acetazolamide administration to augment brain perfusion and reveal areas of diminished reserve *(15–17)*. However, all existing imaging modalities share several common limitations: they cannot be performed under physiological stimuli, akin to exercise in cardiac stress testing; they all require expensive equipment; and their lack of scalability and logistic limitations do not permit implementation at bedside, in the clinic, or widespread use for screening in the community settings. Each method also has its specific limitations. For instance, the Acetazolamide challenge has various side effects and contraindications in the presence of certain comorbidities, while the reference measurements used in CT perfusion assessment are derived from a single unaffected major cerebral vessel in each individual, which in many cases may lead to the underestimation of cerebral blood flow (CBF) in affected regions *(16)*. The Framingham Stroke Risk Profile has recently been augmented with multimodal magnetic resonance imaging (MRI) to quantify the risk of acute stroke in patients with symptomatic ischemia*(18, 19)*. This approach requires contrast injection and access to MRI centers, rendering it impractical for routine stroke risk screening. Further, with the advent of ischemic symptoms, primary prevention likely becomes moot, as the presence of symptoms suggests the progression of vascular disease to a clinically significant stage. There have been efforts using transcranial Doppler (TCD) ultrasonography to evaluate brain blood flow*(20–22)* with technician-dependent varying results. In addition, TCD only targets larger arteries and can only be used on limited areas of the skull, most commonly the temporal acoustic window. A cost-effective and scalable stroke risk assessment system simply does not exist. This makes effective long-term prevention of stroke unfeasible - having established through a questionnaire that a patient is at high risk for stroke, a physician has no good way to know whether a patient's risk is increasing or stable, and whether and when a patient needs intervention.

Physiologically, a stroke occurs when the brain's vascular system is compromised, either through a blockade in blood vessels (ischemic stroke) or a rupture of blood vessels (hemorrhagic stroke). These events lead to impaired blood flow and reduced oxygen supply to brain tissues. Risk factors of stroke include age, cardiovascular diseases, high blood pressure, high blood sugar, smoking, and elevated cholesterol*(10)*. These same factors also impact how the brain reacts to oxygen deprivation or carbon dioxide buildup during events like breath-holding. During breath-holding, both cerebral blood volume and cerebral blood flow increase in response to higher carbon dioxide and lower oxygen levels. To mitigate the increased shear pressure of the blood caused by the rising blood flow, the cerebral autoregulation system prompts blood vessel dilation. Such cerebral autoregulation is impaired in diseased atherosclerotic vessels, including in response to other hemodynamic perturbations such as hypotension and upstream stenosis (e.g., carotid stenosis).

The skull blocks most means for direct, non-invasive measurements of the brain's blood dynamics. Fortunately, infrared light can transmit relatively well through the skull and brain*(23–26)*. By transmitting infrared light through one location on the skull and collecting its transmission from another location, it is possible to track the brain blood volume rate by measuring the light attenuation rate*(27–29)*. If the light used is coherent, such as a laser, mutual interference of light following different trajectories causes speckle patterns, which depend on the coherence of the laser source *(30–33)*. By observing how fast the transmitted laser speckle field fluctuates, it is possible to also determine brain blood flow rate by using a camera to *(27, 34–37)*. This technique is commonly called speckle contrast optical spectroscopy (SCOS)*(27, 34, 35, 38)* or speckle visibility spectroscopy (SVS)*(25, 34, 36, 39, 39, 40)*. The movement of blood within the brain causes this speckle field to fluctuate – the faster the blood flow, the faster the fluctuation. Compared to diffusing wave spectroscopy which utilizes a photodetector and requires a sampling frequency in the MHz range, SCOS which utilizes a CMOS sensor is not only more cost-effective to implement but also enhances signal sensitivity by leveraging the large pixel counts of the camera*(36, 39, 40)*. We previously demonstrated the use of a compact SCOS device on the frontal area of the head to non-invasively measure brain blood flow, by using a large source-to-detector S-D > 3 cm distance*(36)*.

In this paper, we combine a comprehensive brain blood flow and blood volume compact SCOS system with a breath-holding exercise to provide a direct physiology-based measurement of the brain blood vessel condition. Such a system consists solely of a laser diode and a CMOS-based board camera that can be placed on the head with no external optical elements, maintaining a lightweight, portable, and budget-friendly design*(36)*. The monitoring of the regional brain blood flow and volume rate changes during breath-holding was performed on a cohort of 50 subjects, divided into two groups of 25 subjects each, using the Cleveland Clinic Stroke Risk Calculator where the lower the score, the lower the risk*(9)*. One group was at low risk for stroke (low-risk group, score 1), and the other was at higher risk for stroke (higher-risk group, score ≥ 4). Our breath-holding data revealed a statistically significant difference between the two groups when comparing changes in either blood flow or blood volume ($p < 0.05$). Notably, individuals at low risk for stroke showed a higher increase in brain blood volume, while subjects at higher risk for stroke exhibited a greater rise in brain blood flow rates. By comparing the ratio of blood flow changes over blood volume changes—a proxy to blood pressure changes—we observed a statistically significant difference between the groups ($p < 0.0001$)*(27, 41)*. These findings align with our expectations: during breath-holding among those at elevated stroke risk, less flexible blood vessels impede dilation (lack of compliance), prompting accelerated blood flow in response to the brain's heightened demand for oxygen caused by carbon dioxide buildup. We further stratified the higher-risk cohort into smaller sub-groups according to the Cleveland Clinic Stroke Risk Calculator. Our analysis revealed a significant correlations between stroke risk scores and blood pressure ratio during breath-holding. This study offers preliminary physiological insights into stroke risk assessment by establishing correlations between these two methods. We also like to note that SCOS is different from transcranial doppler ultrasound (TCD) *(20–22)*. TCD requires instrument placement over major blood vessels as it depends on unidirectional blood flow that induces a Doppler shift to make measurements. In contrast, SCOS is sensitive to all blood movements, including in blood capillaries, and is therefore capable of more generalized cerebral blood flow measurements.

**Results**

Our wearable SCOS system is shown in Fig. 1A. It includes a laser source for illumination and a board camera for detection. See Methods and Ref. *(36)* for a more detailed presentation of the apparatus. The source injects coherent laser light of wavelength λ = 785 nm into the head of the participant. At a source-to-detector (S-D) distance from the illumination spot, a high-speed camera (60 frames per second (fps)) captured the light that traveled through the brain. The S-D distance was adjusted between S-D = 3.2 cm and 4.0 cm to maintain optimal brain sensitivity and signal-to-noise ratio. It is important to bear in mind that scalp and skull thickness can vary among individuals, which may lead to variations in the ideal S-D distance from one subject to another*(25)*. The brain blood volume was extracted from the camera images by calculating the blood volume index (BVI) from the recorded images as *(27)*:

$$BVI(t) = \frac{I_0}{<I(t)>}, \tag{1}$$

where in Eq. (1) $I_0$ is the intensity at baseline and $I(t)$ is the intensity recorded by the camera image at time t. The intensity baseline is typically measured with the subject at rest, averaged over multiple cardiac cycles. The brain blood flow was extracted from the camera images by calculating the blood flow index (BFI) as*(35, 36)*:

$$BFI(t) = \frac{1}{K^2_{adjusted}(t)}, \tag{2}$$

where in Eq. (2) $K^2_{adjusted}$ is the adjusted speckle contrast obtained by subtracting the raw speckle contrast $K^2_{raw}$ by all sources of noise*(36)*. The raw speckle contrast is defined as:

$$K_{raw}^2(t) = \frac{\sigma^2(I(t))}{\mu^2(I(t))}, \tag{3}$$

where in Eq. (3) $\sigma^2(I)$ is the variance and $\mu(I)$ the mean of the recorded image I at time t. See Refs. *(35, 36)* for more details about speckle contrast and BFI calculations from the recorded camera images. In all the results of this paper, we utilize the normalized BFI and BVI relative metrics for enhanced comparability across measurements. The BFI accounts for the volume of blood moving per unit of time while the BVI accounts for the volume of blood present at the given time. The ratio between BFI and BVI serves as a proxy for relative mean arterial blood pressure measurement*(41, 42)*. Any alterations in the blood pressure can be attributed to changes in the blood flow or changes in the diameter of the blood vessel (i.e. blood volume), which can both be measured with the compact SCOS device.

A typical cerebral blood flow, blood volume, and heart rate measurements recorded during voluntary breath-holding by our compact SCOS device is shown in Fig. 1B. The breath-holding protocol was as follows: the total duration was 180 seconds: 60 seconds of normal respiration, followed by voluntary breath-holding with a self-selected duration as tolerable by the subject, then a recovery period and normal respiration until the 180$^{th}$ second. The participants were asked to exhale before holding their breath and were reminded not to push their limits or hold their breath beyond their capability. The participants were also asked to remain steady and to avoid head movement during the measurement. All the measurements were conducted on hairless areas, i.e. on the forehead region (Fig. 1A).

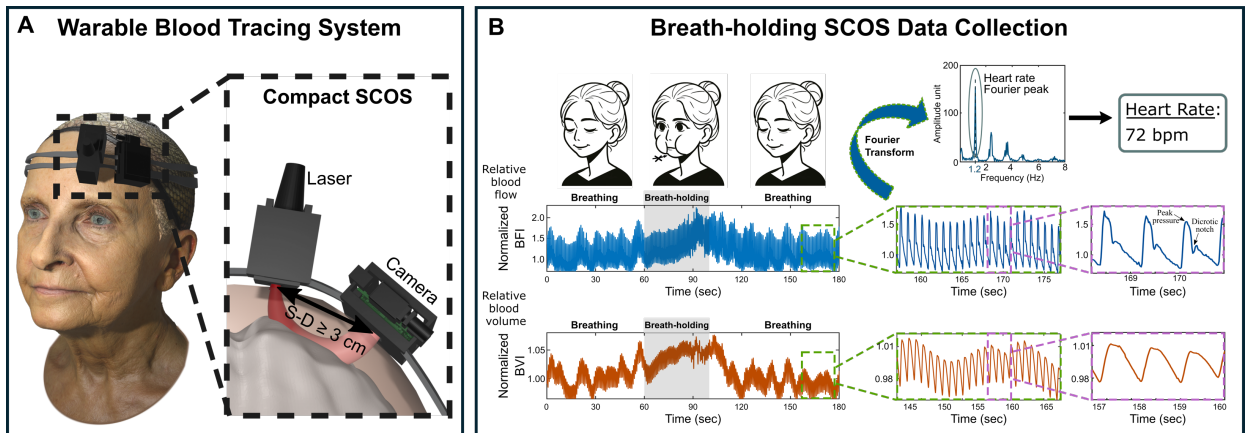

**Fig. 1**: **Brain blood tracing during breath-holding via infrared laser device.** (**A**) Compact speckle contract optical spectroscopy (SCOS) device for blood flow and blood volume simultaneous measurements. (**B**) Breath-holding blood flow and blood volume data collection. The heart rate is extracted from the blood flow via Fourier transform.

A marked increase of the BFI and BVI can be observed during the participant's breath retention as shown in Fig. 1B, attributed to the increased brain's demand for blood to transport oxygen and carbon dioxide until breath retention stops. During breath retention, the brain enters a heightened state of alert, triggering a sequence of protective mechanisms to ensure a stable regulation of carbon dioxide and oxygen*(27, 43, 44)*. This maintenance mechanism is accomplished by mobilizing and diverting oxygen and nutrients from other parts of the body to the brain, achieved through an accelerated circulation of blood and thus an increase in blood flow*(45)* together with an increase in blood volume in the brain through dilation of blood vessels*(46)*. Consequently, a swift and notable increase in BFI and BVI traces can be observed, albeit with some fluctuation. Notably, over a few seconds after the termination of the breath retention, the cerebrovascular reactivity still exhibits a reactive increase in BFI and BVI before returning to the baseline level. Note that the large fluctuations in the BFI and BVI traces in Fig. 1B are not noise but represent blood pulsations as shown in the insets. These pulsations are observable in both BFI and BVI traces with similar overall shape and frequency. However, the vascular waves such as the dicrotic notch and peak pressure are only discernable in the BFI trace (Fig. 1B). This underscores the significance of separately measuring blood flow and blood volume, as they each convey distinct information. The heart rate of the participant can also be extracted by Fourier transforming the BFI signal and measuring the frequency of the heart rate

Fourier peak *(25, 36)*. In Fig. 1B, the heart Fourier peak was measured at a frequency of 1.2 Hz corresponding to a heart rate of 72 bpm (beat-per-minute).

Using the BFI, BVI, and heart rate brain blood tracing during breath-holding, we aim to assess differences between subjects at low risk for stroke and subjects at higher risk for stroke (Fig. 2). To determine the stroke risk of a subject, we used the traditional stroke risk assessment method, shown in Fig. 2A, which is based on the Cleveland Clinic Stroke Risk Calculator*(9)*. This questionnaire relies on indirect markers for stroke risk estimation, such as demographics and comorbidities including age, smoking and low physical activity lifestyle-related factors, diabetes, hypertension, along with previous history of stroke or TIA. This widely recognized questionnaire is used to estimate an individual's risk of having a stroke within the next ten years. Based on the questionnaire input, a 10-year stroke risk probability ranging between 0.7% and 13.4% is computed. The probability score is divided into a scale risk from 1 to 10 where the lower the scale, the lower the risk. A score of $1^{st}$ decile corresponds to a low risk (0.7% probability) of having a stroke in the next ten years. See Supplementary Material for more information about the Cleveland stroke risk questionnaire.

For this study, we enrolled 50 participants divided equally into two groups of 25 subjects each by using the Cleveland Clinic Stroke Risk Calculator. The cohort consisted of subjects with diverse ages, genders, races, and varying levels of stroke risk factors, see Materials and Methods for more details. This study received approval from the Caltech Committee for the Protection of Human Subjects and the Institutional Review Board (IRB). One group was at low risk for stroke (the low-risk group) and corresponded to participants who scored $1^{st}$ decile (0.7% probability). Typically, participants in the low-risk group were individuals younger than 40 years of age and in good health. The second group was at higher risk for stroke (higher-risk group) and corresponded to participants who scored in the $4^{th}$ decile or greater (3% probability or higher). Typically, participants in the higher-risk group were individuals older than 55 years of age and with at least one of the following: history of stroke/TIA, high blood pressure, high blood sugar or diabetes, smoking, history of cardiovascular events, or poor health. Participants with a stroke risk score of 2 or 3 were excluded from this study. The participant with the highest stroke risk score in this study had a score of 7. We recruited 50 participants to ensure sufficient study power.

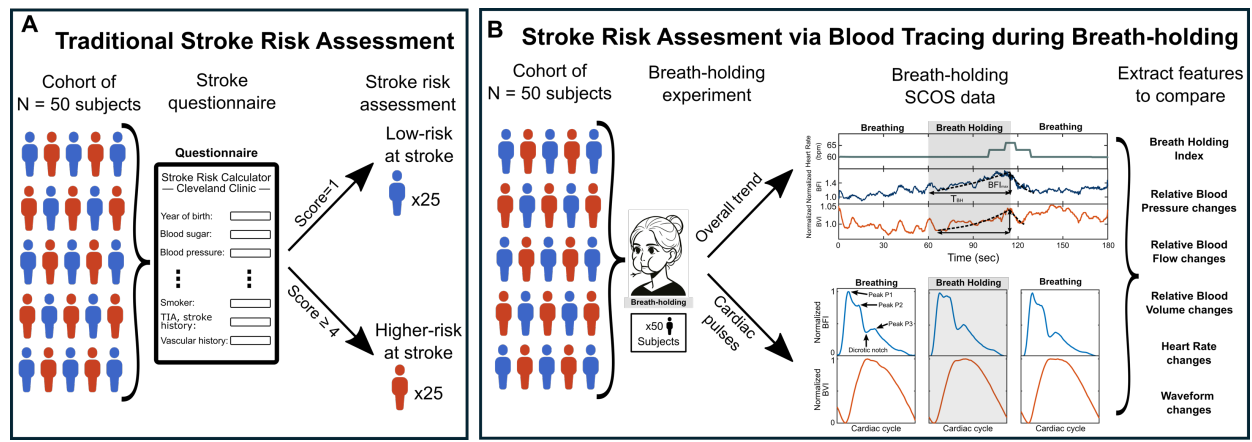

**Fig. 2**: **Stroke risk assessment via blood tracing during breath-holding.** (**A**) Traditional stroke risk assessment based on a questionnaire. (**B**) Proposed stroke risk assessment via blood tracing during breath-holding.

Figure 2B shows the proposed stroke risk assessment via blood tracing during breath-holding (as in Fig. 1B) with the same cohort of participants as in Fig. 2A. A typical example of the heart rate, BFI, and BVI time traces are depicted in Fig. 2B. The heart rate was measured by Fourier transforming a sliding window of 20 seconds in the BFI time trace. The heart rate increase during breath-holding aligns well with the BFI and BVI increases. For the best visualization and extraction of features from BFI and BVI time traces, the temporal resolution was smoothened by a two-second temporal averaging filter. From those, we extracted different features outlined below to compare across the two groups (Fig. 2). Our expectation is to observe a statistically significant difference between the two groups (low-risk vs higher-risk) and correlate back our results with the risk probabilities estimated by the Cleveland Clinic stroke risk calculator, indicating that it is possible to estimate the risk of having a stroke in near future via brain blood tracing during breath-holding.

The first set of features was the breath-holding index (BHI) defined as the maximal change from baseline value during breath-holding divided by the duration of breath-holding $T_{BH}$ *(27, 47)*. We defined a breath-holding index (BHI) for BFI, BVI, and blood pressure ratio as:

$$BHI_F = 100 \cdot \frac{BFI_{max} - BFI_0}{BFI_0 \cdot T_{BH}}, \tag{4a}$$

$$BHI_V = 100 \cdot \frac{BVI_{max} - BVI_0}{BVI_0 \cdot T_{BH}}, \tag{4b}$$

$$\text{Blood pressure ratio} = \frac{BHI_F}{BHI_V}, \tag{4c}$$

where $BFI_{max}$, $BVI_{max}$ denote the maximum value of BFI, BVI during breath-holding, $BFI_0$, $BVI_0$ are the BFI, BVI baseline value before breath-holding when the subject is at rest, and $T_{BH}$ is the duration (seconds) of breath-holding. The $BHI_F$ and $BHI_V$ assess the cerebrovascular reactivity by measuring the percentage change per second of $BFI$ and $BVI$ and have a unit of percent per second*(47, 48)*. The blood pressure ratio in Eq. (4c) serves as an estimate of changes in mean arterial pressure.

The second set of features was extracted by fitting the BFI (or BVI) changes during and after breath-holding using exponential functions:

$$f_1 = 1 + BFI_{max} e^{\frac{T_i - T_{start}}{\tau_{growth}}}, \tag{5a}$$

$$f_2 = 1 + BFI_{max} e^{\frac{T_j - T_{max}}{\tau_{decay}}}, \tag{5b}$$

where $BFI_{max}$ is the maximum value of BFI during breath-holding, $T_{max}$ is the time after breath-holding at which BFI is maximal (i.e. right before BFI starts decreasing), $T_i$ are times from the start of breath-holding until $T_{max}$ and $T_j$ are times from $T_{max}$ until the BFI returns to the baseline level, $\tau_{growth}$ is the growth factor and $\tau_{decay}$ is the decay factor. Both $\tau_{growth}$ and $\tau_{decay}$ are measured from the fits $f_1$ and $f_2$. Our expectation is that the low-risk group exhibit different $\tau_{growth}$ and $\tau_{decay}$ factors than the higher-risk group. The time $T_{max}$ typically corresponds to the sum of $T_{start}$ (time at which breath-holding starts) plus $T_{BH}$ (the duration of breath-holding). The resting heart rate and maximal heart rate changes during breath-holding were also compared between the two groups, see Supplementary Material.

The BFI and BVI time traces were also segmented by each cardiac cycle (Fig. 2B) and the waveforms before, during, and after breath-holding were compared. As shown in Fig. 2B, the typical BFI cardiac pulse is composed of three distinct peaks P1, P2, P3, and one dip (dicrotic notch), related to arterial waveform. Although our method for measuring blood flow is based on a different principle than arterial pressure measurements, we observed similarities between our measurements and arterial pressure map tracings found in the literature*(49)*. Specifically, peaks P1, P2, and P3 can be associated with changes in arterial pressures during the cardiac cycle. The first peak P1 corresponds to the rapid ejection of blood (systole). The second peak P2 corresponds reflected waves from the vascular tree*(49)*. After the ventricular ejection, the pressure within the ventricles drops further marking the end of systole, producing a dip in the BFI, i.e. dicrotic notch (beginning of diastole). The third peak P3 corresponds to the closing of the aortic valve preventing backflow into the heart, causing a brief momentary rise in the BFI (diastole). Interruption of the diastolic downstroke caused by P3 produces the dicrotic notch. It has been shown that these peaks may help quantify aortic stiffness and cardiovascular health condition*(50)*. Furthermore, as shown in Fig. 2B, the waveform of the BFI cardiac cycle, specifically the relative height of the P1, P2, and P3 peaks changes before and during breath-holding, as observed in Ref. *(27)*. Note that the peaks P1, P2, P3, and the dicrotic notch are not present in the BVI cardiac cycle. To quantitatively evaluate the relative height of the P1 and P2 peaks, we calculated the cardiac peaks height ratio $peaks\ ratio = BFI_{P2}/BFI_{P1}$ from the normalized BFI pulses. As reported in Ref. *(27)*, we found that the ratio increased during breath-holding.

Figure 3 summarizes the results from the breath-holding experiment conducted on the cohort of 50 participants. For each participant, between two to five datasets of breath-holding were recorded, depending on the subject's performance in the breathing exercise. From each dataset (such as the one shown in Fig.

2B), features such as BHI (Eqs. 4a-4c), $\tau_{growth}$ (Eq. 5a), $\tau_{decay}$ (Eq. 5b), $T_{max}$ and $T_{BH}$ were extracted. Then, for each subject, the extracted features were averaged across all realizations. The averaged features for each subject are presented in Fig. 3. Subjects were categorized into two groups: low-risk and higher-risk groups. As shown, statistically significant differences can be observed between the low-risk group and the higher-risk group. Figure 3A shows the typical BFI and BVI breath-holding time traces of each risk group, obtained by averaging the features in the $f_1$ and $f_2$ fit across all subjects of each risk group. As shown, the higher-risk group exhibits a higher and slightly faster increase in blood flow during breath-holding than the low-risk group. On the contrary, the higher-risk group exhibits a lower and slightly slower increase of blood volume during breath-holding. Note that both risk groups have similar $T_{max}$, i.e. similar duration of breath-holding $T_{BH}$. These findings align with our expectations: during breath-holding among those at elevated stroke risk, less flexible blood vessels impede dilation, prompting accelerated blood flow in response to the brain's heightened demand for oxygen caused by carbon dioxide buildup. As a consequence, reactive blood pressure increases more for the higher-risk group than the low-risk group due to the shear effect and lower capability of the older population to dilate vessels, indicative of lower vascular compliance.

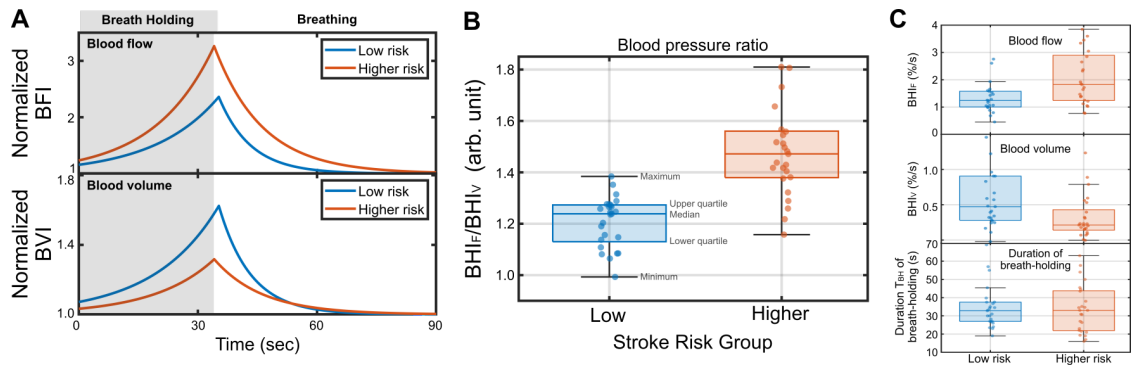

**Fig. 3**: **Breath-holding features comparison between the low-risk and higher-risk groups.** (**A**). Illustrative breath-holding changes in BFI and BVI during breath-holding for each group. (**B**). Blood pressure ratio breath-hold index comparison. (**C**). Relative blood flow and blood volume breath-hold index comparison. Breath-holding duration comparison. See Table 1 for p-values and significance levels.

Those results are further validated in Fig. 3B by looking at the distribution of the blood pressure ratio $BHI_F/BHI_V$ defined in Eq. (4c). In Figs. 3B-C, the data is presented using a box plot, where each data point corresponds to each subject. The top and bottom edges of the box correspond to the upper and lower quartiles, the line inside the box corresponds to the median and the outside lines correspond to the maximum and minimum in the data set excluding any outliers. Outliers are data points that significantly differ from the rest of the dataset and are defined as data points whose values are larger than 1.5 times the interquartile of the top or bottom of the box. As expected, the median and quartiles of the blood pressure ratio are higher for the higher-risk group than the low-risk group, with a statistically significant difference (p-value < 0.0001) as shown in Table 1. Figure 3C shows the BHI distributions for the BFI and BVI. We also show the distribution of the duration of breath-holding $T_{BH}$ for both groups in Fig. 3C, where no statistical difference between the two groups was observed, indicating that both groups perform similar breath-holding exercise. We also compared other features such as the resting heart rate (p-value = 0.467) and maximal heart rate (p-value = 0.005) between the two groups, see Table 1 and Supplementary Material. We also compared and correlated the results of Fig. 3 to those obtained using a blood pressure cuff located on the upper arm, see Supplementary Material.

    The mean and standard deviation of the features are presented in Table 1, along with the corresponding p-value and statistical significance. As shown in the table, the breath-holding duration $T_{BH}$ was similar across both groups. There is also no significant difference in the time constants $\tau_{growth}$ and $\tau_{decay}$ between both groups. Notably, the higher-risk group exhibited, on average, greater flow changes but smaller volume changes during breath-holding compared to the low-risk group. We attribute this difference to the likelihood that individuals in the higher-risk group may have lower vascular elasticity than those in the low-risk group. Note that the BHI ratio $BHI_F/BHI_V$ exhibits a more pronounced difference

between the two groups, as shown by a lower p-value (<0.0001) than the $BHI_F$ or $BHI_V$ feature. Additionally, the higher-risk group has a higher standard deviation for $BHI_F/BHI_V$ than the low-risk group, which we attribute to a more diverse range of health conditions within this group. Results indicate that the blood pressure ratio ($BHI_F/BHI_V$) is the most robust discriminator among the features listed in Table 1.

| Feature | Low-Risk Group mean (std) | Higher-Risk Group mean (std) | p-value | Significance level |
|---|---|---|---|---|
| Stoke risk score | 1 (0) | 5 (1) | | |
| $T_{BH}$ (s) | 35 (12) | 34 (13) | 0.797 | ns |
| $\tau_{growth}$ (s) | 16.3 (8.6) | 15.0 (8.3) | 0.593 | ns |
| $\tau_{decay}$ (s) | 8.0 (6.1) | 12.1 (17.5) | 0.280 | ns |
| $BFI_{change}$ (%) | 44 (16) | 63 (25) | 0.003 | ** |
| $BVI_{change}$ (%) | 20 (14) | 9 (8) | 0.001 | ** |
| $BHI_F$ (%/s) | 1.36 (0.57) | 2.26 (1.62) | 0.014 | * |
| $BHI_V$ (%/s) | 0.62 (0.49) | 0.32 (0.30) | 0.011 | * |
| $BHI_F/BHI_V$ (arb. u.) | 1.21 (0.10) | 1.50 (0.23) | 0.000001 | **** |
| Resting heart rate (bpm) | 68 (9) | 66 (8) | 0.467 | ns |
| Maximum heart rate (bpm) | 84 (10) | 76 (10) | 0.005 | ** |
| $T_{BVI}^{max} - T_{BFI}^{max}$ (s) | 1.4 (2.3) | 1.0 (2.6) | 0.601 | ns |
| Peaks ratio$_{resting}$ (arb. u.) | 0.84 (0.13) | 1.05 (0.23) | 0.003 | ** |
| Peaks ratio$_{BH}$ (arb. u.) | 1.04 (0.14) | 1.22 (0.23) | 0.009 | ** |
| Peaks ratio$_{BH}$/Peaks ratio$_{resting}$ | 1.24 (0.08) | 1.18 (0.11) | 0.0517 | ns |

**Table 1: Mean and standard deviation of the different features extracted from the breath-holding dataset, comparing the low-risk and higher-risk groups.** P-values were calculated from the Welch's t-test. The significance levels are indicated by stars as where ns: not significant, *: p < 0.05, **: p < 0.01, ***: p < 0.001, and ****: p < 0.0001.

We also analyzed the changes in the vascular peaks height ratio ($peaks\ ratio = BFI_{P2}/BFI_{P1}$) between peaks P1 and P2 (Fig. 2B), which was found to increase during breath-holding*(27)*. For the peak analysis, data from nine subjects were removed due to low quality, resulting in a cohort size of 41 (17 low-risk and 24 higher-risk groups). We initially segmented the BFI time trace into individual cardiac pulse and applied a min-max normalization to ensure each pulse's BFI was comprised between 0 to 1 (Fig. 2B). Subsequently, we measured the relative heights of the two peaks P1 and P2 (see Fig. 2B), and calculated their ratio. We then compared the peaks ratio at rest ($peaks\ ratio_{resting}$) and during breath-holding ($peaks\ ratio_{BH}$) between the two risk groups. As shown in Table 1, statistically significant differences between the two risk groups were observed, with the peak ratio for the higher-risk group being higher than that for low-risk group, as similarly observed in Ref. *(50)* by recording Doppler waveforms in 286 participants on the carotid artery. Finally, we compared the ratio between $peaks\ ratio_{BH}/peaks\ ratio_{resting}$ but could not find any statistically significant differences among the two risk groups. Overall, these results indicate that our device can be a valuable tool for assessing cerebralvascular health conditions. Its high sampling rate allows for the acquisition of many cardiac pulses during a single breath-holding sequence, ideal for input into deep learning methods that require larger sample sizes. See Supplementary Material for additional discussion on other features comparisons between the two groups.

Finally, we subdivided the distribution of $BHI_F/BHI_V$ within the higher-risk group into risk score groups of 4, 5, and 6-7, using the Cleveland Score Risk score. The low-risk group (score risk of 1) had a sample size of $N_1 = 25$ subjects, while the higher-score risk subgroups 4, 5, and 6-7 had a sample size of $N_4 = 6$, $N_5 = 10$, and $N_{6-7} = 9$ subjects, respectively. Despite the limited sample sizes hindering statistically significant conclusions, the data in Figure 4 shows an increasing trend in the $BHI_F/BHI_V$ ratio together with the risk score, showing a correlation with the Cleveland Stroke Risk Calculator.

Notably, Fig. 4 includes one instance of an abnormally high blood pressure ratio $BHI_F/BHI_V$ (labeled Outlier) in the risk group 6-7. We anticipate that this subject may have a potentially higher risk of stroke compared to the rest of the cohort, however, this conclusion remains tentative due to the small sample size.

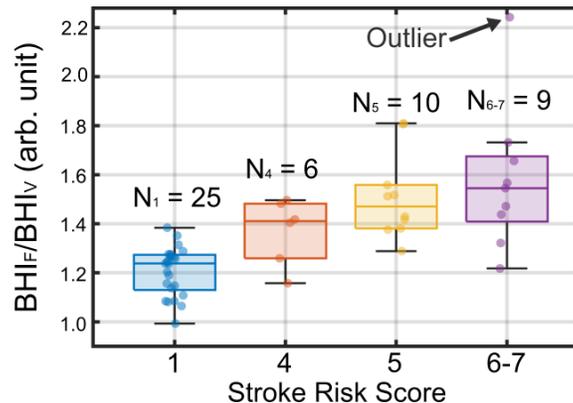

**Fig. 4:** Changes of $BHI_F/BHI_V$ ratio during breath-holding exercise for groups of risk scores 1, 4, 5, and 6-7.

**Discussion**

We combined a compact and cost-effective laser-powered device (compact SCOS) with a breath-holding exercise for assessing cerebrovascular reactivity via measurement of blood blow and blood volume changes on a cohort of 50 participants stratified into two groups (low-risk vs higher-risk for stroke). The compact SCOS device consisted of only two components: a laser diode and a CMOS-based board camera*(36)*. The collected breath-holding data showed a statistically significant difference (p-value <0.0001) between the two groups when comparing the $BHI_F/BHI_V$ ratios. The higher-risk group exhibited a higher increase in brain blood volume (p-value = 0.011) and a lower increase in blood flow during breath-holding (p-value = 0.014). The results also correlated well with those obtained using a blood pressure cuff located on the upper arm. These findings align with our expectations that during breath-holding among those at elevated stroke risk, less flexible blood vessels impede dilation, prompting accelerated blood flow in response to the brain's heightened demand for oxygen and clearance of carbon dioxide buildup. In the future, by characterizing the dynamic brain responses to breath-holding, including the extent of changes in BFI and BVI and the speed at which the brain returns to a baseline level of activity, we aim to objectively quantify the cerebrovascular health of a participant and evaluate their individualized risk of experiencing a cerebrovascular disease such as stroke. Due to the portable and cost-effectiveness of the compact SCOS device, we would like to underscore the potential of deploying the device and incorporating the breath-holding maneuver in various geographical settings to provide quantitative feedback for stroke pre-diagnosis and prevention, especially in resource-limited populations which are disproportionately more affected by stroke and its sequelae.Compared to other methods for measuring cerebral blood changes, the approach of simultaneously measuring both blood flow and volume provides additional information for potential stroke evaluation. This dual measurement allows the monitoring of cerebral perfusion pressure*(41, 51, 52)*. Moving forward, we aim to expand our sample size and incorporate machine learning to enhance the processing and analysis of the data, allowing for more detailed and quantitative categorization of stroke risk. With a larger cohort, we could potentially enhance stroke risk assessment through machine learning models trained on subjects' data with a 2 to 5 years follow-up after the experiment to determine if they experienced a stroke. A broader dataset across stroke risk scores would also enable the determination of an optimal threshold for further longitudinal studies. Such a threshold could be derived from the Receiver Operating Characteristic (ROC) curve, where subjects exceeding this threshold could be monitored more closely over extended periods to better understand the progression of their health conditions. Additionally, a larger cohort would improve our ability to conduct controlled comparisons that account for age and other factors, further confirming the observed correlation between measured metrics like $BHI_F/BHI_V$ ratio and stroke risk.

**Materials and Methods**

A. Participants

Participants for this study were recruited from the Caltech and Pasadena/Los Angeles communities, selected among adult humans aged from 18 to 65 years. Subjects uncomfortable with holding their breath or with major respiratory diseases were excluded from this study. Before the experiments, each participant completed a health questionnaire (showed in Supplementary Material), and their blood pressure was recorded. Informed consent was obtained from the parrticipants beforehand. The human research protocol for this study received approval from the Caltech Committee for the Protection of Human Subjects and the Institutional Review Board (IRB). To simplify the experiment and implementation of the device, measurements were conducted on hairless areas, such as the forehead. The total illumination power was within the American National Standards Institute (ANSI) laser safety standards for maximum skin exposure of a 785 nm laser beam.

A total of fifty-three subjects were enrolled. However, three subjects were excluded from the analysis due to poor data quality or significant movement during the recording. The final dataset encompasses 160 breath-holding entries from 50 subjects. According to the prior questionnaire, the mean(std) ages of the low and higher risk groups were 31(5) and 60(4) years old respectively, 33 subjects were Female and 17 were Male. Traditional stroke risk assessment was performed using the Cleveland Stroke Risk Calculator, which provides a probability stroke risk score ranging from 1 to 10, with lower scores indicating lower risk. Among the 50 subjects, 25 scored 1, 6 scored 4, 10 scored 5, and 9 scored between 6 and 7 in the Cleveland Stroke Risk Calculator.

B. Compact SCOS device

The experimental configuration of the SCOS device is shown in Fig. 1A. For this study, we used a single-mode continuous wave 785 nm laser diode [Thorlabs L785H1] which can deliver up to 200 mW. To ensure control over the illumination spot size and prevent undesirable laser light reflections, we housed the laser diode within a 3D-printed mount*(36)*. The laser source and detecting units were positioned on one side of the participants' foreheads (see Fig. 1). The forehead was chosen as the device location, as SCOS devices works better on hairless areas. The laser diode was set 5 mm away from the skin of participants such that the illumination spot diameter was 5 mm. The total illumination power was limited to 45 mW to ensure that the laser light intensity level of the area of illumination is well within the American National Standards Institute (ANSI) laser safety standards for maximum permissible exposure (2.95 mW/mm$^2$) for skin exposure to a laser beam at 785 nm*(53)*. We used a USB-board camera [Basler daA1920-160um (sensor Sony IMX392)]. For optimal performance and stability, we typically operated the camera at a framerate of 60 frame-per-second (fps) and with a global shutter setting recording 12-bit images. The camera featured a pixel pitch of 3.4 µm, which offers a balance between the average intensity per pixel and the number of speckles per pixel*(54)*.

**Funding**
This research was supported by the National Institutes of Health — Award No. 5R21EY033086-02 and the USC Neurorestoration Center.
**Acknowledgments**
The authors thank Maya Dickson for her assistance during the design of the SCOS device.

**Author contributions:**
Conceptualization: CL, CY
Methodology: YXH, SM, AA, JMT, YTL, PDL, JR, CL, CY
Investigation: YXH, SM,
Visualization: YXH, SM, AA, YTL, JR, CL
Funding acquisition: CL, CY

Project administration: SM, AA, CL, CY
Supervision: CL, CY
Writing – original draft: YXH, SM, AA, JMT, YTL, PDL, JR, CL, CY
Writing – review & editing: YXH, SM, AA, JMT, YTL, PDL, JR, CL, CY

**Competing Interests**
The authors declare that they have no competing interests.

**Data Availability**
The data that support the findings of this study are available from the corresponding author upon reasonable request.

**Supplementary**
A Supplementary Material document containing additional supporting results and explanations is attached to this submission.